\documentclass[12pt]{iopart}
\usepackage{citesort}

\newcommand{\beq}{\begin{eqnarray}}
\newcommand{\eeq}{\end{eqnarray}}

\usepackage[]{graphicx}
\begin{document}

\title[How fast can quantum annealers count?]{How fast can quantum annealers count?}

\author{Itay Hen}
\address{Department of Physics, University of California, Santa Cruz, California 95064, USA}
\ead{itayhe@physics.ucsc.edu}

\begin{abstract}
We outline an algorithm for the Quantum Counting problem using Adiabatic Quantum Computation (AQC).
We show that using local adiabatic evolution, a process in which the adiabatic procedure is performed at a variable rate, the problem is solved with 
the same complexity as 
the analogous circuit-based algorithm, i.e., quadratically faster than the corresponding classical algorithm.
The above algorithm provides further evidence for the potentially powerful capabilities of AQC as a paradigm for more efficient problem solving on a quantum computer, and may be used as the basis for solving more sophisticated problems.   
\end{abstract}

%\keywords{Adiabatic quantum computing, Quantum annealers, Grover's search algorithm, Quantum counting} 

\maketitle                  

\section{\label{intro}Introduction}

The paradigm of Adiabatic Quantum Computation (AQC), proposed by 
Farhi et al.~\cite{farhi_long:01} about a decade ago is a simple yet intriguing approach to problem solving on a quantum computer. 
Unlike the leading paradigm of circuit-based quantum computing, AQC is an analog, time-continuous method which does not require the design and use of quantum gates.
As such, it can in many ways be thought of as a simpler and perhaps more profound method for performing quantum computations while also being 
easier to design and implement experimentally~\cite{vandersypen:01,johnson:11,gaitan:12,berkley:13}.

The potential advantages of AQC over the usual circuit-based and other computational paradigms
have generated substantial research,
and intensive efforts have been made towards finding computational problems that are solved faster
on an adiabatic quantum computer (henceforth, a quantum annealer) than on a classical computer (see, e.g.,~\cite{hogg:03,farhi:02,farhi:08,altshuler:09,knysh:11,young:08,young:10,hen:11,hen:12} and references therein).
One of the most important results in that context, which establishes the universality
and quantum power of the AQC paradigm is the existence of a theorem 
stating that adiabatic quantum computing is polynomially equivalent to circuit model quantum computation~\cite{aharonov:07,mizel:07}. 

Despite the above equivalence, it is not yet understood 
in which cases adiabatic quantum computation is as efficient as circuit model quantum computation (or even potentially polynomially faster). 
It would therefore be interesting to look at problems for which circuit model algorithms are known, and to see whether 
equally-efficient adiabatic algorithms can be found. 
To date, however, there are almost no examples to show that AQC algorithms are similar in complexity to circuit-model algorithms. 
One notable exception is the adiabatic Grover's search algorithm devised by Roland and Cerf~\cite{roland:02}, demonstrating 
that the quadratic speedup over classical algorithms first shown by Grover~\cite{grover:97} using circuit-based computation, can also be gained 
with adiabatic quantum algorithms (we discuss this algorithm in more detail later). 

In what follows, we consider another problem for which, much like Grover's search problem, there is a circuit-model algorithm 
that is quadratically faster than its classical counterpart. This is the problem of `Quantum Counting'~\cite{brassard:98,nielsen:00}
in which one has to estimate the number of occurrences 
of a specific item in an unsorted database. We show that the Quantum Counting problem may also be solved on a quantum annealer
quadratically faster than the best corresponding classical algorithms via the use of local adiabatic evolution~\cite{roland:02}. 
%by repeatedly running the QAA with Grover-type adiabatic profiles and measuring the energy of the system at the end of each run. 
We shall see that the complexity of the proposed method is on par with the analogous circuit-based algorithm and so is quadratically faster than the corresponding classical algorithm.

The paper is organized as follows. In the next section we briefly discuss the principles of the Quantum Adiabatic Algorithm 
that is the heart of AQC. We then describe in some detail the concept of local adiabatic evolution in section~\ref{sec:lae}. 
In section~\ref{sec:agp}, we analyze the dynamical equations of the QAA for the adiabatic Grover's algorithm.
These are later used in sections \ref{sec:qc1} and \ref{sec:qc2} to devise a quantum-adiabatic algorithm for solving the problem of Quantum Counting with quadratic speedup. Finally, in section~\ref{sec:sac}, we conclude with a few comments.

\section{\label{qaa}Quantum Adiabatic Algorithm}

We first describe the Quantum Adiabatic Algorithm (QAA) which provides the general approach for solving problems by casting them into the form of optimization problems~\cite{farhi_long:01}.  

Within the framework of the QAA approach,
 the solution to an optimization problem is encoded in the ground state of a Hamiltonian
$\hat{H}_p$. 
To find the solution, the QAA prescribes the following. As a first
step, the system is prepared in the ground state of another Hamiltonian
$\hat{H}_d$, commonly referred to as the driver Hamiltonian.  The driver
Hamiltonian is chosen such that it does not commute with the problem
Hamiltonian and has a ground state that is fairly easy to prepare. 
As a next step,
the Hamiltonian of the system is slowly modified from $\hat{H}_d$ to
$\hat{H}_p$, using the linear interpolation, i.e.,
\begin{equation}
\hat{H}(s)=s \hat{H}_p +(1-s) \hat{H}_d \,,
\end{equation}
where $s(t)$ is a parameter varying smoothly with time,
from $0$ at $t=0$ to $1$ at the end of the algorithm,
at $t=\mathcal{T}$.  If this process is done slowly enough, the
adiabatic theorem of Quantum
Mechanics (see, e.g.,~\cite{kato:51} and~\cite{messiah:62})
ensures that the system will stay close to the ground state of the
instantaneous Hamiltonian throughout the evolution, so that the system evolves to a state that is close to the ground state of $\hat{H}_p$.  At this point,
measuring the state will give the solution of the original problem with high
probability. 

In the simple case where $s(t)$ varies from zero to one at a constant rate,  
the runtime $\mathcal{T}$ must be chosen to be large enough so that the adiabatic
approximation holds: this condition determines the
efficiency, or complexity, of the QAA.  A condition\footnote{
We note that the conditions given here are in general neither 
necessary nor sufficient. More rigorous conditions may be found in, e.g.,~\cite{tong:05,schaller:06,jansen:07}. 
The `traditional' condition stated here is however sufficient for adiabatic search algorithms such as the one discussed here~\cite{jansen:07}.
} on $\mathcal{T}$ can
be given in terms of the instantaneous eigenstates $\{ | m
\rangle \}$ and eigenvalues $\{E_m \}$ of the Hamiltonian $H(s)$,
as~\cite{wannier:65,farhi:02}
\begin{equation} \label{eq:T}
\mathcal{T} >\frac1{\epsilon} \, {\max_{s} V_{01}(s)  \over
\min_s g^2(s)} \,,
\end{equation} 
where $\epsilon$ is a small number, inversely proportional to the running time of the algorithm, that controls the expected probability for failure of the algorithm, $g(s)$ is the first
excitation gap $E_1(s)-E_0(s)$ and $V_{01}(s) = \left| \langle 0 | \rmd H / \rmd s | 1\rangle \right|$ (in our units $\hbar=1$).
\section{\label{sec:lae}Local Adiabatic Evolution}

In the framework of Local Adiabatic Evolution (LAE), the adiabatic parameter $s(t)$ is varied not at a constant rate 
but rather at a variable rate, slowing down in the vicinity of the minimum gap and speeding-up in places where the gap is large~\cite{roland:02}.
The procedure is based on formulating a `local' Landau-Zener condition for each value of the adiabatic 
parameter $s$.  This condition can be written as: 
\beq
\left| \frac{\rmd s }{\rmd t} \right| \leq \epsilon  \frac{g^2(s)}{V_{01}(s)} \,,
\eeq
where 
$g(s)$, $V_{01}(s)$ and $\epsilon$ are as in (\ref{eq:T}). 

When applicable, LAE can provide a significant speedup in calculation times~\cite{roland:02,roland:03,hen:13b}. 
A well-known example to demonstrate this, was first considered by Roland and Cerf~\cite{roland:02}.
In this example, an adiabatic algorithm for solving Grover's search problem~\cite{grover:97} which describes a search for a marked item in an unstructured database, was found. More specifically, Roland and Cerf showed that while the application of the
adiabatic theorem with a linear $s(t)$ results in a running time that is of order $N$, where $N$ is the number of items in the database, 
a carefully chosen variable rate of $s(t)$ yields a running time that scales with $\sqrt{N}$, i.e., a quadratic speedup is gained, similarly to the 
original result by Grover, that was found for the circuit-based model. 

LAE however can only be efficiently used in cases where one has proper knowledge of the exact behavior of the gap and relevant matrix elements of the system 
for the problem at hand. This is normally not the case. 
The amenability of Grover's search problem to LAE stems from the fact that the gap of the system (as well as the relevant matrix element)
can be calculated explicitly beforehand, even if the specific ground state is not known. 
With a proper choice of problem and driver Hamiltonians (which are discussed next), the gap in the Grover search problem is given by the simple expression:
\beq \label{eq:gap}
g(\eta;s)=\left[ 1-4(1-\eta)s(1-s) \right]^{1/2} \,,
\eeq 
where $\eta=M/N$ is the ratio of the number of solutions $M$ to the size of the database $N$.
Moreover, the matrix element $\left| \left \langle \rmd \hat{H} / \rmd s \right \rangle \right|$
can also be calculated in a straightforward manner to give:
\beq
V_{01}(s)=\left| \left \langle 0 \left| \rmd \hat{H} / \rmd s \right| 1 \right \rangle \right| = \frac{\sqrt{\eta(1-\eta)}}{g(\eta; s)}
\eeq

The condition for local adiabatic evolution for the Grover problem is thus simply given by:
\beq \label{eq:lae}
\left| \frac{\rmd s }{\rmd t} \right| \leq \epsilon \frac{g^{3}(\eta; s)}{\sqrt{\eta(1-\eta)}} \,,
\eeq
A less strict inequality, using the fact that $V_{01}(s)<1$, yields the more convenient bound of~\cite{roland:02}:
\beq \label{eq:lae2}
\left| \frac{\rmd s }{\rmd t} \right| \leq \epsilon g^{2}(\eta; s) \,.
\eeq

Integrating in time either of the two equations above between $t=0$ (also, $s=0$) and $t=\mathcal{T}$ (s=1) yields in the limit of $N \to \infty$:
\beq \label{eq:runtime}
\mathcal{T} \propto \frac1{\epsilon} \sqrt{N/M} = \frac1{\epsilon \sqrt{\eta}}
\eeq
i.e., a quadratic speedup over the corresponding classical algorithms, a result also shared by Grover's original circuit-based algorithm~\cite{grover:97}. 

\section{\label{sec:agp}Analysis of the adiabatic Grover problem}

An important property of the gap of Grover's search problem, (\ref{eq:gap}), may be formulated as follows.
\beq 
g(\eta_1;s) \leq g(\eta_2;s)  \quad 
\textrm{for any}  \quad \eta_1 \leq \eta_2 \,,
\eeq
where $\eta_1$ and $\eta_2$ are two arbitrary ratios. 
The above property leads to the interesting (and somewhat surprising) attractive feature of LAE for the Grover search problem:
If one runs the algorithm as if the ratio of solutions is $\eta_*$  even though the actual ratio of solutions $\eta$ may not be known, 
then one still obtains at the end of the adiabatic run a state that is very close the symmetric superposition of all solution states, as long as $\eta_* \leq \eta$.
This is because choosing the adiabatic profile with $\eta_*$, namely,
\beq \label{eq:ds2dtetaStar}
\frac{\rmd s}{\rmd t} = \epsilon g^2(\eta_*; s) \,,
\eeq  
also satisfies the inequality~(\ref{eq:lae2}) for any $\eta_* \leq \eta$.

The above feature is to be contrasted with the circuit-based Grover's search algorithm, in which case one must know in advance the number of solutions 
in order to obtain a solution to the problem with a very high probability.
%This feature is to be contrasted with the analogous circuit-based algorithm (i.e., Grover's algorithm)
%in which case one must know beforehand the number of solutions $M$ (equivalently, the ratio $\eta$) in order to obtain the correct solution
%for the Grover problem with high probability. 
Within the circuit model, the problem of not knowing the number of solutions is resolved  by a preliminary application of a
different algorithm known as Quantum Counting, designed to estimate the number of solutions $M$ prior to the application of Grover's search algorithm. 

In what follows we show that LAE, with the aid of the unique properties of the adiabatic Grover's problem presented above,
may be used to solve adiabatically the problem of Quantum Counting, i.e., the problem of estimating the number $M$ of occurrences of a specific item within an unsorted database of size $N$. 

Before doing so however, we first analyze in some detail 
the dynamics of the adiabatic algorithm for Grover's search problem (a similar derivation is briefly discussed in~\cite{roland:02}).  
Specifically, we derive here the equations describing an unsorted-database search with an unknown number (equivalently, ratio) of solutions $M$ (or $\eta$) using an adiabatic profile $s(t)$ with a ratio $\eta_*$. We then calculate the probability of a solution state at $s=1$, in two limiting cases.

\subsection{Dynamics of the Grover-adiabatic problem}

Consider the problem Hamiltonian for the Grover problem~\cite{roland:02}:
\beq \label{eq:hp}
\hat{H}_p =1 - \sum_{m \in \mathcal{M}} | m\rangle \langle m | \,.
\eeq
Here the $M$ states $m \in \mathcal{M}$ are eigenstates of the computational basis with eigenvalue zero (all other states have eigenvalue one).
% [the driver Hamiltonian is sufficiently implementable]
The problem Hamiltonian is therefore a diagonal matrix of size $N$ with $M$ zero entries  and $(N-M)$ entries of one along the diagonal.
The driver Hamiltonian is simply given by:
%~\footnote{We note here that the one-dimensional projection is efficiently implementable, as it is an operator 
%that distinguishes one specific $n$-bit state from all the rest, where $n=\log_2 N$. This operator can be represented as a product
%on $n$ one-spin operators. See also, e.g.,~\cite{viamontes:05}.}:
\beq
\hat{H}_d = -  | \phi \rangle \langle \phi | \,,
\eeq
where $|\phi \rangle$ is the fully-symmetric state:
\beq
| \phi \rangle = \frac1{\sqrt{N}} \sum_{i=1}^N | i \rangle \,. 
\eeq
Obviously, the ground-state of the driver Hamiltonian, i.e., the state that the system is initially prepared in, is 
\beq
|\psi(t=0) \rangle = | \phi \rangle \,.
\eeq
The Schr\"odinger equation
\beq
\rmi \frac{\rmd}{\rmd t} |\psi(t)\rangle =\hat{H} |\psi(t) \rangle
\eeq
with $| \psi(t) \rangle=\left(c_0(t), c_1(t),\ldots, c_i(t)), \ldots, c_N(t) \right)$ simplifies considerably due to symmetry considerations: 
The state of the system has only two distinct amplitudes. 
The amplitudes $c_i(t)$ that correspond to `solution' states and those that
correspond to `non-solution' states. Let us denote the former set by $a(t)$ and that of the latter by $b(t)$.
The Schr\"odinger equation thus becomes the two coupled equations:
\numparts
\beq
\rmi \dot{a} &=  -(1-s) \left[\eta a+(1-\eta)b\right] & \\ 
\rmi \dot{b} &=  -(1-s) \left[\eta a+(1-\eta)b\right] &+s b \,,
\eeq
\endnumparts
Note that $N$, the size of the problem and $M$, the number of solutions appear in the equations  only as the ratio $\eta=M/N$.

The time-dependence of $a$ and $b$ is easily transformed 
into $s$ dependence (as there is a one-to-one correspondence between the two in our case). 
The time-derivative transforms in the usual way: $\rmd/\rmd t = \left( \rmd s/\rmd t \right) \times \rmd/\rmd s$,
and the two coupled equations now become:
\numparts
\beq \label{eq:coupled}
\rmi \epsilon g^2(\eta_*;s) a' &= -(1-s) \left[\eta a+(1-\eta)b\right] & \\ 
\rmi \epsilon g^2(\eta_*;s) b' &= -(1-s) \left[\eta a+(1-\eta)b\right] & +s b\,.
\eeq
\endnumparts 

Here, the prime symbol $(')$ stands for differentiation with respect to the new independent variable $s$
and we have used~(\ref{eq:ds2dtetaStar}) for $\rmd s/ \rmd t$ 
where the ratio $\eta_*$ may in general be different than $\eta$.
As already noted earlier, a choice of $\eta_* \leq \eta$ results in a true adiabatic evolution and is expected to yield
at the end of the QAA, a state that is very close to the equal superpositions of all solution states. 

In addition, the adiabatic process dictates the initial conditions
\beq a(s=0)=b(s=0)=1\,,
\eeq
which correspond to the fully-symmetric state (i.e., an equal superposition of all possible computational-basis states).
The specific value of $1$ chosen here leads to the simple normalization condition
\beq
\eta |a|^2+ (1-\eta) |b|^2 = 1 \,,
\eeq
that, similarly to the equations of motion, does not explicitly depend on $N$ or $M$ but only on their ratio.

\subsection{End probabilities}
 
At the end of the evolution, at $s=1$, the final state of the system is a linear combination
of the equal superposition of solutions states, namely, of $|\psi_{\textrm{sol}} \rangle  \propto  \sum_{m \in \mathcal{M} } | m \rangle$ 
and the equal superposition of the non-solution states $|\psi_{\textrm{no-sol}} \rangle  \propto  \sum_{m \notin \mathcal{M} } | m \rangle$.
The closer the evolution is to being truly adiabatic, the larger the amplitude of the solution-states superposition will be. 
While the exact values of the two complex components at $s=1$ can not be computed analytically,
one can obtain an asymptotic solution in two important cases. One is the limit of small $\epsilon$, i.e., when the allowed error in the final state goes to zero. 
In the $\epsilon \to 0$ limit, the two equations are of the WKB type~\cite{bender:78}, and have a series solution of the form
\beq
a(s)=\e^{\frac{\rmi}{\epsilon} \sigma(s)}\,
\textrm{with} \, \sigma (s)=\sum_{k=0}^{\infty} (\rmi \epsilon)^k \alpha_k(s)
\eeq
which can directly be plugged into the equations [note that $b(s)$ is a simple linear combination of $a(s)$ and $a'(s)$]. 
Solving the resulting equations for the growing powers of $\epsilon$ term by term, one may calculate the
probability of obtaining a solution state, i.e., the squared overlap between the final state at $s=1$ and the optimal solution $|\psi_{\textrm{sol}} \rangle$:
\beq
P_{\textrm{sol}} &=& |\langle \psi(s=1) | \psi_{\textrm{sol}} \rangle |^2 \\\nonumber
&=& \lim_{s \to 1}  \frac{\eta |a(s)|^2}{\eta |a(s)|^2+(1-\eta) |b(s)|^2} \,.
\eeq
Calculation of the above series solution leads to the following asymptotic solution for $P_{\textrm{sol}}$ in the limit of small $\epsilon$:
\beq \label{eq:Plimit}
P_{\textrm{sol}}(\epsilon \to 0)  = 1-\epsilon^2 \eta (1-\eta) +O(\epsilon^4) \,.
\eeq
To leading order in $\epsilon$, the probability to find a solution does not depend on value of $\eta_*$,
as long as $\epsilon$ is small enough. As expected, when $\epsilon$ goes to $0$ (i.e., when no error is allowed), the probability of the system to be in the equal superposition of solution states at the end of the QAA approaches $1$ (in this case, the running time of the algorithm goes to infinity). 
%$\epsilon^4 \eta (1-\eta) ( 4-77 \eta +105 \eta^2 +64 \eta_*-112 \eta \eta_* +16 \eta_*^2)$.

The more interesting limit however, is the one in which $\epsilon$ is small but kept fixed while $N$, the dimension of the Hilbert space (alternatively, the size of the database)
goes to infinity. This limit translates to the limit where the ratios $\eta=M/N$ and $\eta_*=M_*/N$  are small, i.e., $M, M_* \ll N$~\footnote{We note that in the case where the number of solutions $M$ is of order $N$, the Quantum Counting problem becomes less interesting, as the classical probability for finding a solution, $P_{\textrm{sol}}^{\textrm{(classical)}}=M/N$, becomes significantly large, in which case no quantum speedup
is gained.}. In this limit, we obtain, perhaps expectedly, the following `Landau-Zener type' probability:
\beq \label{eq:Plimit2}
P_{\textrm{sol}}(N \to \infty) \approx 1-\e^{-\frac{\pi \eta}{4 \epsilon \eta_*}} = 1 - \e^{-\frac{\pi M}{4 \epsilon M_* }}\,.
\eeq
As expected, the probability of obtaining a solution 
increases with decreasing $\epsilon$ and also, as discussed in the previous section, with decreasing $M_*$. Most importantly, in the $N \to \infty$ limit, the above probability has a well established $N$-independent value. The above probability will become useful in what follows when we address the problem of Quantum Counting.

\section{\label{sec:qc1}Quantum counting: Zero vs nonzero number of solutions}

Before addressing the full-complexity of the general quantum counting problem, let us first 
consider  the special case of finding whether or not the number of solutions $M$ (number of zero entries in the problem Hamiltonian) is zero.
Circuit-based algorithms give a running time that is on the order of $O(\sqrt{N})$~\cite{brassard:98,nielsen:00}, that is, a quadratic speedup
over the analogous classical algorithm.

To solve this problem quantum-adiabatically, one executes the QAA algorithm with the adiabatic profile function $s(t)$ that obeys 
\hbox{$\rmd s / \rmd t = \epsilon g^2(\eta=1/N,s)$}, i.e., a profile 
that is suitable for solving the case of any nonzero number of solutions with quantum speedup.

Now, if the number of solutions $M$ is zero, then the problem Hamiltonian will have no effect on the state of the system,
and the state will remain the initial state $| \phi \rangle$ throughout the adiabatic evolution, except for an undetectable global phase. 
A measurement of the $z$-magnetization at the end of the run will produce a random configuration with energy $1$.

If on the other hand the number of solutions is nonzero, the Grover LAE procedure 
ensures us that at the end of each of the run, at $s=1$, the final state of the system will be very close to $|\psi_{\textrm{sol}} \rangle$, the equal superposition of zero-energy states with a probability greater than $P_{\textrm{sol}} = 1 - \e^{-\frac{\pi M}{4 \epsilon}}$ (in the $N \to \infty$ limit), and a measurement along the $z$ direction will produce a zero-energy state with this probability.
 
We conclude then, that one execution of the adiabatic algorithm with runtime $\mathcal{T} \propto \sqrt{N}/\epsilon$ is sufficient 
for establishing whether or not the number of solutions is zero. 

\section{\label{sec:qc2}Quantum counting: The general case}
The general case of estimating the number of solutions $M$ can be analyzed with the aid of the probability $P_{\textrm{sol}}$ obtained in~(\ref{eq:Plimit2}).
Let us first briefly analyze the classical analogue of that same problem. In the classical case, random sampling of items in the database
will give $P_{\textrm{sol}}^{\textrm{(classical)}}=M/N=\eta\,$ for each picked item. The number of successes after $k$ trials will therefore be distributed binomially with the unbiased mean
of $\hat{\eta}^{\textrm{(classical)}} = \eta$ and a sampling error of:
\beq
\Delta \hat{\eta}^{\textrm{(classical)}} \sim \sqrt{\frac{\eta(1-\eta)}{k}} \approx \sqrt{\frac{\eta}{k}} \,.
\eeq
%where the approximation above is due to the small $\eta$ limit. 
In terms of the number of solutions $M$, the error in $M$ is thus
\beq
\Delta \hat{M}^{\textrm{(classical)}} \sim \sqrt{\frac{M N}{k}} \,.
\eeq
For the error to be proportional to $\sqrt{M}$, the number of trials $k$ required for the search, has to scale linearly with the dimension of the database $N$.

In the quantum case, the question thus becomes: How many independent runs $k$ of the adiabatic procedure, each run producing a solution-state with probability $P_{\textrm{sol}}$ [as given by~(\ref{eq:Plimit2})], are necessary 
to yield an unbiased estimate of $M$ with an error that, similarly to the classical case, scales with $\sqrt{M}$?
We answer this question in two steps: First, we find a suitable value of $M_*$, according to which the adiabatic paths for subsequent QAA runs
will be constructed. As a second step, we calculate how many QAA runs with that chosen value of $M_*$ are required to estimate $M$ for a given error rate. 

Note that an optimal value of $M_*$ should be on the order of the unknown $M$. Choosing a value for $M_*$ that is too large would correspond to probabilities that are very close to zero, leading to very poor statistics that would in turn not suffice for the determination of $M$ in an accurate manner. Choosing a value for $M_*$ that is too small would lead to unnecessarily-long running times for the adiabatic runs. An optimal choice for $M_*$ would thus be one that would yield 
a significantly-large probability for obtaining a solution state (one reasonable choice could be, e.g., $P_{\textrm{sol}} \approx 0.1$),
which implies $M_* \sim M/\epsilon$.

Since $M$ is unknown, a way to find an optimal value of $M_*$ that would yield a given probability, is to employ a simple noisy binary search algorithm~\cite{karp:07,boyer:98,krovi:10}. Starting with geometrically diminishing values of $M_*$, namely, \hbox{$M_*=N/2, N/4, \ldots$}, 
the QAA is executed multiple times for each value, such that for each tested value $M_*$, the average ratio of solutions is computed
(note that since $M \ll N$ the ratios will be zero at first). When $M_*$ is in the vicinity of $M$ and the obtained ratios of solutions are
no longer strictly zero, the trial $M_*$ values are adjusted (i.e., increased or decreased) accordingly until an approximate value of the chosen probability of solutions, $P_{\textrm{sol}}$ is obtained. 
Note that in order to obtain a certain probability of solutions $P_{\textrm{sol}}$ with a square-root error accuracy, only a fixed  (i.e., an $M$-independent) number of QAA executions is required for each tested value of $M_*$ (as the number of `solution' outcomes is distributed binomially). For a binary search such as the one outlined above, the number of trial $M_*$ values required is logarithmic in $N$, where the complexity of each trial run is $O(\sqrt{N/M_*})$. The overall complexity of this first step is therefore simply $O(\sqrt{N/M})$.

Having found an optimal value for $M_*$, the second step of the algorithm is initiated, 
consisting of the execution of multiple QAA runs with the chosen value of $M_*$ (note that the chosen value obeys \hbox{$M_* \sim M/\epsilon$}). 
A straightforward calculation analogous to the one performed for the classical case above yields an error of:
\beq
\Delta \hat{M} \sim \frac{4 \epsilon M_*}{\pi} \left( \frac{\e^{\frac{\pi M}{4 \epsilon M_*}}-1}{k}\right)^{1/2}\,.
\eeq
The number of trials needed to obtain an error that scales with $\sqrt{M}$ is therefore:
\beq
k \propto \frac{16 \epsilon^2 M_*^2 \left(\e^{\frac{\pi M}{4 \epsilon M_*}}-1\right)}{ M \pi^2} 
\approx \frac{4 \epsilon M_*}{\pi } \,,
\eeq
where the last approximation is justified by the smallness of the probability $P_{\textrm{sol}}$. 
Since the complexity of each QAA run is $O(\sqrt{N/M_*})$, the overall complexity of the algorithm will be 
\hbox{$O(M_*\sqrt{N/M_*}) = O(\sqrt{N M})$}, similarly to the circuit-based quantum counting algorithm~\cite{nielsen:00}. 
An analogous calculation in which the error is required to scale linearly with the number of solutions, i.e., \hbox{$\Delta M \sim M$}, 
yields a complexity of \hbox{$O(\sqrt{N/M_*}) = O(\sqrt{N/M})$}, also in accord with the circuit-based algorithm~\cite{brassard:98}.

\section{\label{sec:sac}Summary and conclusions}
We have given a prescription for solving the Quantum Counting problem in which the number of occurrences of a specific item 
within an unsorted database is to be estimated. We have shown that using local adiabatic evolution 
on Grover-type problems, the algorithm is quadratically faster than the corresponding classical algorithm and on par with the circuit-based result.

We note here, that similarly to the original Grover's algorithm, the problem Hamiltonian of the Quantum Counting algorithm presented here, 
is given as an oracle, and the quadratic speedup exhibited here holds only relative to it. 

It would be interesting to see whether the algorithm outlined above can be used as the basis for 
the solution to the more general problem of Quantum Phase Estimation~\cite{nielsen:00} 
within which one is required to estimate the eigenphases (i.e., the phases of the eigenvalues) of a given unitary matrix.  
Note that the (unnormalized) fully-symmetric initial wave function $| \phi \rangle$ may be re-expressed in the form~\cite{nielsen:00}:
\beq
| \phi \rangle = \sqrt{\eta} | \alpha \rangle + \sqrt{1-\eta} | \beta \rangle \,,
\eeq
where we have defined:
\beq
| \alpha \rangle &=& \frac1{\sqrt{N \eta}} \sum_{m \in \mathcal{M}} | m \rangle \,,\\
| \beta \rangle &=& \frac1{\sqrt{N(1-\eta)}} \sum_{m \notin \mathcal{M}}  | m \rangle \,.
\eeq
Replacing $\eta=M/N$ with the appropriate angle of $\sin^2 \left( \theta /2\right)$
reveals that estimating $M$ is the same as estimating the angle $\theta$ for the wave function: 
\beq
| \phi \rangle =\sin \left( \theta /2\right) | \alpha \rangle + \cos \left( \theta /2\right) | \beta \rangle \,,
\eeq
meaning that quantum counting may be thought of as a phase estimation algorithm, where the `phase' here is represented 
by the angle $\theta$.

The importance of the above results, aside from the implications resulting from having a quadratically-faster quantum-adiabatic algorithm for this specific problem, lies in the fact that it is one of very few examples that illustrate the power and potential encompassed in adiabatic quantum computation, providing an adiabatic quantum algorithm that is as efficient as its circuit-model counterpart. 

%\begin{acknowledgement}
\ack 
We thank Peter Young, Eleanor Rieffel and J\'er\'emie Roland for useful comments and discussions. 
We acknowledge partial support by the National Security Agency (NSA)
under Army Research Office (ARO) contract number W911NF-09-1-0391, and in part
by the National Science Foundation under Grant No.~DMR-0906366. 
%\end{acknowledgement}

\section*{References}
\bibliography{refs}

%\appendix
%\section*{Appendix}

\end{document}